\documentstyle[12pt,aaspp]{article}
\def\ltsima{$\; \buildrel < \over \sim \;$}
\def\lsim{\lower.5ex\hbox{\ltsima}}
\def\gtsima{$\; \buildrel > \over \sim \;$}
\def\gsim{\lower.5ex\hbox{\gtsima}}
\def\cs{{c_{\rm s}}}
\def\Cs{{c_{\rm s}}}

\def\Ms{{M_{\odot}}}

\def\p{{\partial}}

\def\vp{{V_{\varphi}}}
\def\vr{{V_r}}

\begin{document}
\title{SELF-SIMILAR COLLAPSE OF A SELF-GRAVITATING VISCOUS DISK}

\author{Shin Mineshige\altaffilmark{1}}
\affil{Department of Astronomy, Kyoto University, Sakyo-ku, 
 Kyoto 606-01, Japan}

\and

\author{Masayuki Umemura\altaffilmark{2}} 
\affil{Center for Computational Physics, University of Tsukuba,
Tsukuba, Ibaraki 305, Japan}

\altaffiltext{1}{e-mail: minesige@kusastro.kyoto-u.ac.jp}

\altaffiltext{2}{e-mail: umemura@rccp.tsukuba.ac.jp}

\begin{abstract}
A self-similar solution for time evolution of isothermal, self-gravitating
viscous disks is found under the condition that $\alpha' \equiv \alpha (H/r)$
is constant in space (where $\alpha$ is the viscosity parameter and $H/r$ is
the ratio of a half-thickness to radius of the disk).  This solution describes
a homologous collapse of a disk via self-gravity and viscosity.  The disk
structure and evolution is distinct in the inner and outer parts.  There is
a constant mass inflow in the outer portions so that the disk has flat rotation
velocity, constant accretion velocity, and surface density decreasing outward
as $\Sigma \propto r^{-1}$.  In the inner portions, in contrast, mass is
accumulated near the center owing to the boundary condition of no radial
velocity at the origin, thereby a strong central concentration being produced;
surface density varies as $\Sigma \propto r^{-5/3}$.  Moreover, the transition
radius separating the inner and outer portions increases linearly with time.
The consequence of such a high condensation is briefly discussed in the context
of formation of a quasar black hole.

\end{abstract}

\keywords{accretion, accretion disks --- black hole physics ---
galaxies: kinematics and dynamics --- gravitation --- stars: formation}

\section{INTRODUCTION}

Quasars (QSOs) are the most powerful objects that have ever existed in the
universe.  The emergence of quasars at high-redshifts,
$z \lsim 5 $, is thus crucial when considering
the formation of astrophysical objects, notably of galaxies.
The view is widely accepted that QSO phenomena result from
mass accretion onto supermassive black holes.
However, the formation process of seed black holes
at high redshifts is not well understood at the present.
There are two distinct lines of thoughts concerning this issue.
One is based on considering a formation of a proto-quasar
supermassive black hole
after the formation of a host galaxy
as the consequence of stellar mass loss
and star encounters at the nucleus of the galaxy (Rees 1984).
The other rather assumes a galaxy-independent population of
massive black holes
(Carr, Bond, \& Arnet 1984; Loeb 1993; Fukugita \& Turner 1996).
Under the latter picture, a question is
how quasar black holes formed at high redshifts, $z > 5 - 10$.

Suppose a high density fluctuation with a mass scale of $\sim 10^6\Ms$
began to collapse at high redshifts of $z \gsim 10$.
Such a cloud acquires angular momentum through tidal torque
in the course of a gravitational collapse.
Resultantly formed a rotationally supported, self-gravitating disk.
For a typical spin parameter, the angular momentum barrier is
by roughly seven orders of magnitude larger
than the Schwarzschild radius (Loeb 1993; Eisenstein \& Loeb 1995).
The problem is then how to get rid of angular momentum from the cloud
so as to form a black hole.
Radiation drag via the cosmic background radiation seems to have been
at work at $z > 100$, but is effective only when the cloud is optically thin
(Loeb 1993; Umemura, Loeb, \& Turner 1993; Tsuribe \& Umemura 1996).
Afterwards,
angular momentum in the cloud could be redistributed
via gravitational torque rising from nonaxisymmetric perturbations
(Paczy\'nski 1978; Lin \& Pringle 1987; Papaloizou \& Lin 1989) and/or
turbulent shear viscosity which
could be associated with magnetic fields (Shakura \& Sunyaev 1973).
It is thus worth investigating how a self-gravitating, viscous disk
evolves in the context of black-hole formation.
Furthermore, this kind of study is of great importance, of course, 
when one investigates physics of galaxy and star formation.

The basic equilibrium structure of accretion disks are now well
understood, as long as we believe the standard model based
on the $\alpha$-viscosity prescription (Shakura \& Sunyaev 1973).
Nevertheless, it is not easy to follow its dynamical evolution,
mainly because the basic equations for the disks are highly nonlinear,
especially when the disk is self-gravitating (e.g. Paczy\'nski 1978;
Fukue \& Sakamoto 1992).
To follow nonlinear evolution of dynamically evolving systems, 
in general, the technique of self-similar analyses is sometimes useful.
Several classes of self-similar disk solutions were known previously
(Pringle 1974; Filipov 1984),
but all of them considered a disk in a fixed, external potential.

We are now concerned with dynamical evolution of a self-gravitating
disk in a time-evolving, self-consistent potential.
As far as steady, nonviscous rotating disks are concerned,
there are plenty of works so far done.
Mestel (1963) was the first to find a simple disk solution,
in which physical quantities are integrated vertically
with respect to the disk equatorial plane.
Hayashi, Narita, \& Miyama (1982) found
two-dimensional, isothermal disk solutions with finite temperature
(see also Toomre 1982 for stellar systems).
Numerical steady solutions are calculated by several groups
(Hachisu, Eriguchi, \& Nomoto 1986;
Bodo \& Curir 1992; Hashimoto, Eriguchi, \& M\"uller 1995).
Recently, we have found
a simple analytical solution for a steady, self-gravitating,
isothermal disk (Mineshige \& Umemura 1996, hereafter Paper I) 
as an extension of Mestel (1963) disk.
However, little study has been done concerning dynamical evolution
of self-gravitating, viscous disks.

We, in the present study, seek for a time-dependent, self-similar
solution for a gravitational collapse of a rotation-supported,
self-gravitating viscous disk.  
When a disk is sufficiently cool, gravitational instability will occur 
(Toomre 1964), providing a source of disk viscosity 
(Paczy\'nski 1978; Lin \& Pringle 1987)
or causing disk fragmentation (e.g., Bodenheimer, Tohline, \& Black 1980).
Several authors thus mainly discussed the consequence of gravitational
instability in the context of fueling to active galactic nuclei
(e.g. Shore \& White 1982; Shlosman \& Begelman 1987;
Shlosman, Frank \& Begelman 1989),
or (multiple) star formation (see Boss 1986; Myhill \& Kaula 1992).
We here adopt a rather distinct approach;
we, in the present study, try to find an analytical solution
for a collapse of rotating, viscous disks,
putting aside for the moment the stability argument.
It might be noted in this context that Shu (1977)
found the self-similar solution for a gravitational collapse of
an isothermal sphere.  
Saigo \& Hanawa (1996) discussed the effects of rotation.
We extend these works so as to incorporate
the effects of mass accretion via viscosity.
We derive self-similar solutions in section 2,
and then discuss the formation of a primordial quasar black hole
in section 3.

\section{SELF SIMILAR, SELF-GRAVITATING DISK}

\subsection{Basic Equations for Self-Similar Variables}

We start with the time-dependent version of 
the height-averaged equations for isothermal accretion disks 
(cf. Honma, Matsumoto \& Kato 1991; Narayan \& Yi 1994);
$$     {\p\Sigma\over\p t} 
             + {1\over r}{\p\over\p r}(r \Sigma \vr) = 0,     \eqno(1)$$
$$     {\p\vr\over\p t} + \vr{\p\vr\over\p r}
             = -{\cs^2\over\rho}{\p\rho\over\p r}
               -{GM_r\over r^2} + {\vp^2\over r},             \eqno(2)$$
$$     {\p(r\vp)\over\p t} + \vr{\p(r\vp)\over\p r}
       = {1\over r\Sigma}{\p\over\p r}
            \left({\nu\Sigma r^3 }{\p\Omega\over\p r}\right).
                                                              \eqno(3)$$
Here, $\Sigma = 2\rho H$ is surface density,
$H$ is half-thickness of the disk, $\Omega = \vp/R$,
$\cs$ is sound velocity (which is constant by assumption), 
$M_r$ is the mass of a disk within a radius $r$,
and we approximated a potential to be $\sim -GM_r/r$.
This is a good approximation if $\Sigma(r)$ profile is
steeper than $1/r$ (see Appendix).
We prescribe kinematic viscosity as
$$  \nu = \alpha\cs H = \alpha ({H\over r})\cs r,             \eqno(4)$$
with $\alpha$ being viscosity parameter,
because we will find later that self-similar 
solutions exist if $\alpha' \equiv \alpha (H/r)$ is constant
in space.  From now on, therefore,
we assume $\alpha'$ (instead of $\alpha$) to be constant.
For vertically self-gravitating disks, $H$ is determined as
$$   H = {\cs \over(4\pi G\rho)^{1/2}} = {\cs^2\over 2\pi G\Sigma}, 
                 \quad 
    \rho = {\Sigma\over 2H}
         = {\pi G\Sigma^2\over\cs^2}.                         \eqno(5)$$
To proceed, it is convenient to rewrite mass conservation (1)
using $M_r(r,t)$; 
$$   {\p M_r\over\p t} + \vr{\p M_r\over\p r} = 0, \quad 
      {\p M_r\over\p r} = 2\pi r\Sigma.                       \eqno(6)$$

Now, we introduce the following self-similar variables (Shu 1977);
$$   x \equiv {r\over \cs t}, \quad
     \Sigma(r,t) = {\cs\over 2\pi G t} \sigma(x), \quad  
     M_r(r,t) = {\cs^3t\over G} m(x),          $$
$$   \rho(r,t) = {\sigma^2(x)\over 4\pi Gt^2}, \quad
       H(r,t) = {\cs t\over \sigma(x)},           \quad
     \vr(r,t) = \cs u(x),                      $$
$$  \vp(r,t) = \cs v(x), \quad 
      j(x) \equiv xv = {1\over\cs^2}{r\vp(r,t)\over t}.       \eqno(7)$$
Note that derivatives are transformed into
$$  {\p\over\p t} \rightarrow -{x\over t}{\p\over\p x}
                              +          {\p\over\p t'},\quad
    {\p\over\p r} \rightarrow  {x\over r}{\p\over\p x},        \eqno(8)$$
for the transformation, $(r,t)\rightarrow(x,t'=t)$.
Since all the time derivatives with respect to $t'$ disappear
if we use self-similar variables (Eq. 7),
we hereafter write $d/dx$ instead of $\p/\p x$.

Equation (6) now becomes
$$   m + (u-x){dm\over dx} = 0, \quad{\rm and}\quad
       {dm\over dx} = x\sigma,                                \eqno(9)$$
yielding a simple relation between $m$, $\sigma$ and $u$;
$m = x \sigma (x - u)$.
With this being kept in mind, equations (1) -- (3) can be modified as
$$   (u-x) {1\over\sigma}{d\sigma\over dx} 
            + {du\over dx} + {u-x\over x} = 0,                \eqno(10)$$
$$   {2\over\sigma}{d\sigma\over dx} + (u-x){du\over dx} 
            - \sigma{u-x\over x} - {v^2\over x} = 0,          \eqno(11)$$
$$ j + (u-x){dj\over dx} 
     = \alpha' {1\over \sigma x}{d\over dx}
       \left[\sigma x\left(-2j + x{dj\over dx}\right)\right]. \eqno(12)$$

\subsection{Solution in a Slow Accretion Limit}

In the limit of slow accretion ($v \gg 1, \sigma \gg 1, |u| \ll 1$),
equation (11) gives
$$  v = \sigma^{1/2}(x-u)^{1/2}, \quad
        j = \sigma^{1/2} x(x-u)^{1/2}.                       \eqno(13)$$
leading to
$$   {d\ln j\over d\ln x} 
         = 1+{1\over 2}{1\over x-u}\left(x-{du\over d\ln x}\right)
                 +{1\over 2}{d\ln\sigma\over d\ln x}.         \eqno(14)$$
Note that from equation (10) we derive
$$   {d\ln\sigma\over d\ln x} 
             = {1\over x-u}{du\over d\ln x} - 1,              \eqno(15)$$
from equation (12).
Inserting equation (15) into equation (14), we have
$$   {d\ln j\over d\ln x} = {1\over 2} + {1\over 2}{x\over x-u}
         = {2x-u\over 2(x-u)}.                                \eqno(16)$$
After some algebra, we obtain
$$   {u\over 2x} 
       = - \alpha' {1\over \sigma x j}{d\over dx}
              \left({\sigma x j}{2x-3u\over x-u} \right).     \eqno(17)$$
With a help of the expressions for $j$ (Eq. 13) and $\sigma$ (Eq. 15), 
we finally derive an ordinary differential equation for $u(x)$:
$$  {du\over dx} = -{4x^2-6ux+3u^2 \over 2(x-3u)x}
                   -{1\over\alpha'}{u(x-u)^2\over (x-3u)x}.  \eqno(18)$$
Equation (18) can easily be integrated numerically
for an appropriate boundary condition; $u=0$ at $x=0$
if we assume no central object (such as a black hole).
Once $u=u(x)$ is obtained, we can derive $\sigma = \sigma(x)$
by integrating equation (15) for a given
$\sigma_0 \equiv \sigma(x=1)$.  The results of the integration
are displayed in figure 1 for different values of
$\alpha' = 10^{-3}, 10^{-2}$, and $10^{-1}$.
The azimuthal velocity is derived from equation (13).

Note that each physical quantity is a rather smooth function of $x$.
We generally find $du/dx < 0$;
that is, $u(x)$ is a monotonically decreasing function of $x = r/\cs t$.
Furthermore, physical quantities, such as $u$ and $\sigma$,
are power-law functions of radius in the limits of 
$x \gg \alpha'$ and $x \ll \alpha'$.

In the limit of large $x (\gg \alpha')$,
mass accretion is induced by viscosity.
Two terms on the right-hand side of equation (18) are balanced 
with each other (while $du/dx = 0$).  
We find
$$   u \approx - 2\alpha',           \quad 
     \sigma \approx \sigma_0 x^{-1}, \quad
     v \approx \sigma_0^{1/2},       \quad
     {\dot m} \approx 2\alpha'\sigma_0,                      \eqno(19)$$
where $\dot m (\equiv -x\sigma u)$ corresponds to a mass-flow rate.
The radial dependences of physical quantities at large $x$
are the same as those of the stationary, self-similar solution 
of a self-gravitating viscous disk (Paper I).
However, we find $\vr \approx -2\alpha\cs(H/r)$ in the current
time-dependent solution, whereas $\vr = -\alpha\cs(H/r)$ in the
steady solution.  This indicates that accretion velocity is doubled
when we consider the effects of continuously growing central mass
(see discussion in Paper I).

In the limit of small $x \ll \alpha'$
the first term dominates over the second on the right-hand side 
of equation (18),
$$   u \approx -2x \left(1 - {9\over 11}{x\over\alpha'}\right), \quad 
     \sigma \approx {\sigma_0\over\alpha'} 
                     \left({x\over\alpha'}\right)^{-5/3}
                     \left(1 + {8\over 11}{x\over\alpha'}\right),    $$
$$   v \approx (3\sigma_0)^{1/2} 
                     \left({x\over\alpha'}\right)^{-1/3}
                     \left(1 + {1\over 11}{x\over\alpha'}\right), \quad
     {\dot m} \approx 2\alpha'\sigma_0 
                     \left({x\over\alpha'}\right)^{1/3}
                     \left(1 - {1\over 11}{x\over\alpha'}\right). \eqno(20)$$
Note that $u$ (and therefore $\vr$) is not proportional to $\alpha'$,
indicating that mass-inflow is not controlled by viscosity,
but is regulated by the inner boundary condition of $\vr = 0$ at
$r = 0$.  Mass is thus being accumulated continuously near the origin.

To sum up, the disk structure and evolution is distinct in
the inner and outer parts.  The transition radius ($r_{\rm tr}$)
separating these two parts increases linearly with time, 
because $r_{\rm tr} \approx \alpha'\cs t \propto t$ 
for a fixed $\alpha'$ (Eq. 7).
We thus assume $r_{\rm tr} = 0$ initially; in other words,
we consider the later evolution of the disk with
$\Sigma \propto r^{-1}$ everywhere.
(This is the situation postulated in Paper I.)
As matter accretes towards the center, 
$\Sigma$ profile changes from inside.

Now we recover physical variables from self-similar ones using
equation (7): we obtain
$$   \vr \approx -2\alpha'\cs, \quad
     \Sigma \approx \Sigma_0 \left({r\over r_0}\right)^{-1},  \quad
     \vp \simeq (2\pi G\Sigma_0r_0)^{1/2}, \quad 
 {\dot M} \simeq 4\pi\alpha'\cs r_0\Sigma_0,                 \eqno(21)$$
at large $r/t$ ($\gg \alpha'\cs$), and
$$   \vr \approx -2\cs \left({r\over r_0}\right)
                       \left({t\over t_0}\right)^{-1},   \quad 
     \Sigma \approx \Sigma_0 
                   \left({r\over r_0}\right)^{-5/3}
                   \left({t\over t_0}\right)^{2/3},  $$
$$   \vp \approx \cs \left({r\over r_0}\right)^{-1/3}
                     \left({t\over t_0}\right)^{1/3},  \quad
     {\dot M} \approx 4\pi r_0\Sigma_0\cs
                     \left({r\over r_0}\right)^{1/3}
                     \left({t\over t_0}\right)^{-1/3},       \eqno(22)$$
at small $r/t$ ($\ll \alpha'\cs$).
Here, $\dot M \equiv -2\pi r\Sigma \vr$ 
is a dimensional mass-flow rate and we approximated
$M_r \approx \int^r 2\pi r_0 \Sigma_0 dr = 2\pi \Sigma_0 r_0^2$ 
in equation (21).  The units are
$$  r_0 = 1.0~r_{\rm pc} ~{\rm pc}, \quad
    \cs \simeq 10^{6.0} T_4^{1/2}~{\rm cm~s}^{-1}, \quad
    t_0 \equiv {r_0\over \cs} \sim 
          10^{5.0}  {r_{\rm pc}\over T_4^{1/2}}~{\rm yr},     $$
$$    {\dot M}_0 \equiv 4\pi\alpha'\cs r_0\Sigma_0
         \sim 10^{0.27}{M_6 T_4^{1/2}\over r_{\rm pc}}~\Ms{\rm ~yr}^{-1},
                                                              \eqno(23)$$
for temperature of $\sim 10^4T_4$K, mass of $\sim 10^6M_6\Ms$, 
respectively.
The unit for $\Sigma$ is chosen so as to give
$M = \int 2\pi \Sigma(r)~rdr$ for the initial state, in which
$\Sigma = \Sigma_0 r_0/r$;
For such normalizations, a normalization constant of $\sigma(x)$ is
$$  \Sigma_0 = {M\over 2\pi r_0^2}
         \sim 10^{1.5}{M_6\over r_{\rm pc}^2}~{\rm g~cm}^{-2},\quad
       \sigma_0 \equiv {2\pi Gt_0\over\cs}\Sigma_0
              \sim 10^{1.71}{M_6\over r_{\rm pc}T_4}.        \eqno(24)$$
Note that $\sigma_0$ represents the ratio of disk radius to height
at $x=1$ (see Eq. 7), or the initial ratio of gravitational energy
to thermal energy of the disk, $\vp^2/\cs^2$ (Eq. 21).
The model parameters of the self-similar
solutions are $\alpha'$, $\cs$ (or temperature), and $\sigma_0$.

Figure 2 plots the time evolution of a self-gravitating disk.
Clearly, there are two regimes as mentioned previously (cf. Fig. 1).
The radius separating the outer and inner parts
is increasing linearly with time.  If we follow a disk evolution at
a fixed $r$, hence, we see that $\vr$ is initially constant
and then decreases at $t > r/\alpha'\cs$.
Accordingly, mass inflow rate also decreases with the time, 
causing a rapid growth of $\Sigma$ and $M_r$.
Note that since $H/r \sim (x\sigma)^{-1}$ (Eq. 7),
$H/r$ is constant at large $r/t$, 
while it rapidly decreases inward; $H/r \propto (r/t)^{2.5}$.
The thin disk and slow accretion approximations
are even better in the inner portions at later times, 
although $\alpha$ may exceed unity at $x \ll \alpha'$.
This means, the present solution does not give
a good representation of the disk structure
at $r/t \ll \alpha'\cs$ (discussed later).

\bigskip
\section{DISCUSSION}
\medskip
\subsection{Summary of the Self-Similar Solution}

We have derived a self-similar solution for time evolution of an 
isothermal, self-gravitating, viscous disk in the slow accretion limit.
Disk structure changes from the inner to outer parts.
For example, surface density is scaled as $r^{-5/3}$ in the inner,
while it is $r^{-1}$ in the outer.  This interface gradually
moves outward in proportion to $t$.
In this solution
density increases monotonically with the time at the center.
The mass profile near the center is
$$    M_r(r) = \int_0^r 2\pi \Sigma(r)~rdr
              \simeq 3\times 10^6 \left({r\over r_0}\right)^{1/3}
                         \left({t\over t_0}\right)^{2/3} 
                           {r_{\rm pc}\over T_4^{1/3}}\Ms.   \eqno(25)$$
[Although this yields a diverging $M_r$,
the increase of $M_r$ should be terminated in a realistic situation,
when the outer disk is depleted with gas.]

As claimed first by Mestel (1963) and also by Paper I,
the thin-disk approximation breaks down at radii
comparable with the thickness.
In fact, the present solution gives diverging $\Omega$ and $\alpha$
as $x$ approaching $0$, which suggests that the solution does not
represent physical situation at $x \ll 1$.
Moreover, since we assume steady mass input towards the center,
the central mass condensation increases at any time.
Once a central object forms from a central mass condensation,
gravity is dominated by this object at sufficiently small radii,
where we may adopt a solution for a point-mass potential.

Realistically, there may be two or three zones in a disk.
Before forming an object, a self-gravitating disk has
two zones (as mentioned in previous section).
After the formation of a central object, in contrast
there are three zones; the innermost region is dominated by a 
point-mass potential and the other two zones are dominated by
self-gravity of the disk.
Since $\dot M > 0$, the mass of the central object
is continuously increasing with time.
The transition radius between the innermost to the inner region again 
increases linearly with time (Paper I).

Self-similar solutions assume that heating and cooling
rates have the same radial dependence (see Eq. 4 in Paper I).
A flat temperature distribution is the result of this assumption.
This is a reasonable approximation at least in the outer regions:
when we balance viscous heating and radiative cooling rates
in a thin-disk approximation, 
we find $\cs \propto r^{-1/12}\sim r^{-3/13}$,
depending on the optical depth of the disk and opacity sources (Paper I).
This relatively flat temperature profile results from 
the fact that for $\Sigma \propto r^{-1}$ (as in the outer parts)
the potential is logarithmic and thus has a weak radial dependence.
At $x \ll 1$, in contrast, this approximation may break down, since
potential has stronger radial dependence.  The isothermal approximation
may not be justified at the innermost region at later times ($r \ll \cs t$).

A self-gravitating disk is locally stable, if
$$   Q \equiv {\Cs\kappa\over \pi G\Sigma}  \gsim 1,            \eqno(26)$$
as long as the effects of viscosity and radial mass inflow are ignored
(Toomre 1964).
Here, $\kappa$ is epicyclic frequency and
$\kappa = 2^{1/2}\Omega$ for $\Omega \propto R^{-1}$.
If we simply apply this criterion to the present model, we find
$Q \simeq 2^{3/2}(H/R)^{1/2}$ at $x \gsim 1$ (Eq. 5 and 21),
indicating that the disk is stable for $H/R \gsim 1/8$.
If $H/R$ is small, gravitational instability will set out,
making disk turbulent, thickening the disk (Paczy\'nski 1978).
However, this is a very naive picture and a more sophisticated
stability analysis, similar to
Christodoulou et al. (1995a, 1995b) but including the effects of disk
viscosity and radial gas inflow, is needed as future work.

\bigskip
\subsection{Formation of a Quasar Black Hole}
\medskip

When $M_r$ exceeds a critical value at some radius,
$$   M_{\rm crit}(r) = (r/10^{5.4}{\rm cm})^{2/3}\Ms,       \eqno(28)$$
the cloud will start to collapse due to a general relativistic 
instability (Shapiro \& Teukolsky 1983), resulting in
the formation of a black hole.
Equation (28) gives a critical mass (for a given
radius) for spherical supermassive stars, while
we are now concerned with evolution of a rotation-supported disk.
Nevertheless,
we employ the argument concerning spherical stars 
in order to see qualitative effects of general relativity,
since a thin-disk approximation breaks down anyway 
near the center as mentioned above,
and since a solid analysis of a collapsing self-gravitating disk
based on the general relativistic formulation
is not available at this moment.

With this being kept in mind,
we discuss a fate of a rotationally supported, viscous disk
with a mass of $\sim 10^6\Ms$, a temperature of $\sim 10^4$K,
and a size of several pc.
In the present picture, such a relatively high disk temperature is
preferable, since otherwise the disk will stay molecular rather than ionized.
The accretion timescale is inversely proportional to the temperature,
and hence it may exceed the age of the Universe
for a molecular disk with $\alpha < 0.01$ 
(e.g. Eq. 1 in Sasaki \& Umemura 1996), 
unless alternative mechanisms,
such as gravitational torque, remove the disk angular momentum.
There are several possibilities to heat the disk.
First, if the formation of primordial hydrogen molecules proceeds
more slowly than the dynamical collapse,
gas will not cool below $\sim 10^4$K.
This may occur if residual free electrons recombine quickly
due to density enhancement, thereby suppressing
the formation of a sufficient amount of H$^-$ ions, 
which help to make hydrogen molecules
(see Hutchins 1976, Palla, Salpeter, \& Stahler 1983).
Second, if the Universe was reionized through first-generation stars
or objects, the disk will be effectively heated by strong UV background
radiation (e.g. Sasaki \& Umemura 1996).
Finally, if star formation occurs within the disk itself,
the disk material can be photoionized by stars.

Figure 3 depicts 
the evolution of such a disk (by the solid lines) and
the critical line for a gravitation instability (by the dotted line)
in the ($\log r$-$\log M_r$) diagram.
As time goes on, the disk becomes more and more condensed at the
center, thereby increasing its mass within a fixed radius.
The mass profile is $M_r \propto r^{1/3}$ (Eq. 25) according 
to the self-similar solution, while
the critical value gives $M_{\rm crit} \propto r^{2/3}$ (Eq. 28).
The solid line should cross the dotted line at
$$   r_{\rm crit} \simeq 10^{11.8}\left({t\over t_0}\right)^2
                           {r_{\rm pc}^3\over T_4^{3/2}}~{\rm cm}, \quad
     M_r(r_{\rm crit}) \simeq 10^{4.3}~\left({t\over t_0}\right)^{4/3}
                           {r_{\rm pc}^2\over T_4}~\Ms.      \eqno(29)$$
We get a condensation of $\sim 10^3\Ms$
on a timescale of $\sim 0.1~t_0 \sim 10^4$yr.

The estimates above are optimistic, however,
since it takes $r_0/\cs \sim 10^5$yr to 
$r_0/(\alpha' \cs) = 10^6(\alpha'/0.1)^{-1}$yr 
for accreting gas to reach the center, and thereby establishing
a self-similar evolution of the disk.  
We thus safely conclude that
within a timescale of $\sim 10^5(\alpha')^{-1}$yr a central region with
a mass of $10^{4-5}\Ms$ could become unstable,
which may give rise to a proto-quasar black hole at high redshifts.
Again, a general relativistic study of a collapsing rotating disk
is necessary to conclude whether this scenario can work or not.

\acknowledgements

We thank the referee for valuable comments 
and T. Hanawa and T. Tsuribe for useful conversation.
This work is partial fulfillment of the Japan-US cooperative research
program which is supported by Japan Science Promoting Foundation and
National Science Foundation on US side.
Analyses were, in part, made at Center for Computational Physics
in University of Tsukuba, and Princeton University Observatory.
Also, this work was supported in part by the Grants-in Aid of the
Ministry of Education, Science, Sports,
and Culture of Japan, 06233101, 08640329 (S.M.) and 06640346 (M.U.).

\appendix
\section{Self-gravity under a thin-disk approximation}

The most straightforward expression for the potentials under
thin-disk approximation is
$$  \Psi(r) = 2G\int_0^\pi d\theta \int_0^{r_0}
                 {\Sigma(R) RdR\over(r^2+R^2-2rR\cos\theta)^{1/2}}, \eqno(A1)$$
(e.g. Mestel 1963), where $r_0$ denotes the size of the disk
and we ignored vertical mass distribution in the disk.
After some algebra, we have
$$    {d\Psi\over dr} = {G(I_1 + I_2 + I_3)},                       \eqno(A2)$$
where $I_1$, $I_2$, and $I_3$ represent
the Keplerian term, finite contributions from the mass within $R$,
and the mass beyond $R$, respectively, and are
$$  I_1 \equiv {1\over r^2}\int_0^r 2\pi R\Sigma(R) dR,     $$
$$  I_2 \equiv 2\pi \sum_{k=1}^\infty\alpha_{2k}\left({(2k+1)\over r^{2k+2}}
                    \int_0^r R^{2k+1}\Sigma(R)dR-\Sigma(r)\right)$$
$$  I_3 \equiv2\pi \sum_{k=1}^\infty\alpha_{2k}\left(\Sigma(r)-
                    2k r^{2k-1}\int_r^{r_0}{\Sigma(R)\over R^{2k}}dR \right),
                                                                   \eqno(A3)$$
with
$$   \alpha_{2k} \equiv {1\over\pi}\int_0^\pi P_{2k}(\cos\theta)d\theta
                   = \left[{(2k)!\over(2^kk!)^2}\right]^2,        \eqno(A4)$$
($P_{2k}$ is the Legendre function; see Eq. 24 of Mestel 1963).
When $\Sigma(r) = \Sigma_0 r_0/r$, in particular, we find
$$   {d\Psi\over dr}
        = {2\pi G\Sigma_0r_0\over r}
             \left[1 + \sum_{k=1}^\infty\alpha_{2k}
                          \left({r\over r_0}\right)^{2k}\right]. \eqno(A5)$$
We, hence, understand that 
if $\Sigma(r)$ profile is steeper than $1/r$
we may approximate gravitational attraction force to be
$-GM_r/r^2$ except near the outer edge.

\clearpage
\begin{figure}
\caption{Radial profiles of the self-similar variable, $\sigma(x)$,
as functions of $x \equiv r/\cs t$.
The three solid lines represent the calculated values of
$\sigma/\sigma_0$ for $\alpha' = 10^{-3}, 10^{-2}$,
and $10^{-1}$, respectively, where $\sigma_0 \equiv \sigma(x=1)$.
The transition radius at $x \sim \alpha'$ separates
the outer part, where $\sigma \propto x^{-1}$, 
and the inner part, where $\sigma \propto x^{-5/3}$, in each curve.
Note that the dotted line corresponds to $\sigma/\sigma_0 = x^{-1}$.
}
\end{figure}

\begin{figure}
\caption{Time evolution of a self-gravitating disk.
>From the top to the bottom, time development of
$\dot M$ distribution,
$\Sigma$ profile, and radial distributions of $\vp$ 
(by the dashed line) and $\vr$ (by the solid line).
The units are $r_0 =$ 1pc, $\dot M_0 \sim$ 2 $\Ms$yr$^{-1}$,
$\Sigma_0 \sim 30$g cm$^{-2}$,
$\cs \sim 10$km s$^{-1}$, and $t_0 \simeq 10^5$yr, respectively.
Parameters are $\alpha' = 0.1$ and $\sigma_0 = 50$.
The elapsed times are $t/t_0 = 0.1$ (indicated by i),
1.0, 10, and $10^2$ (indicated by f), respectively.
}
\end{figure}

\begin{figure}
\caption{Evolution of mass profiles of a self-gravitating disk
with a total mass of $10^6\Ms$, a temperature of $10^4$K,
and a size of 1pc (by the solid lines).  The attached numbers
represent the elapsed times; $t/t_0 = 0.1, 1.0$, and 10.
We assumed $\alpha' = 0.1$ and $\sigma_0 = 50$.
Also displayed are the critical line for a general relativistic
instability, $r_{\rm crit}$ (by the dotted line), 
and the Schwarzschild radius
$r_{\rm g}$ (by the short-dashed line), respectively.}
\end{figure}

\end{document}